\documentstyle[11pt,newpasp,twoside]{article}
\begin{document}
\title{MODEST: modeling stellar evolution and (hydro)dynamics}
\author{Piet Hut}
\affil{Institute for Advanced Study, Princeton, NJ 08540, USA}

\begin{abstract}
Simulations of dense stellar systems currently face two major hurdles,
one astrophysical and one computational.  The astrophysical problem
lies in the fact that several major stages in binary evolution, such
as common envelope evolution, are still poorly understood.  The best we
can do in these cases is to parametrize our ignorance, in a way that
is reminiscent of the introduction of a mixing length to describe
convection in a single star, or an alpha parameter in modeling an
accretion disk.  The hope is that by modeling a whole star cluster in
great detail, and comparing the results to the wealth of observational
data currently available, we will be able to constrain the parameters
that capture the unknown physics.  The computational problem is one of
composition: while we have accurate computer codes for modeling
stellar dynamics, stellar hydrodynamics, and stellar evolution, we
currently have no good way to put all this knowledge together in a
single software environment.  A year ago, a loosely-knit organization
was founded to address these problems, MODEST for MOdeling DEnse
STellar systems, with nine working groups and a series of meetings
that are held every half year.  This report reviews the first year of
this initiative.  Much more detail can be found on the MODEST web site
\ {\tt http://www.manybody.org/modest.html} .
\end{abstract}

\section{Introduction}

Large-scale computer simulations in astrophysics require teamwork.
Gone are the days that an individual graduate student could write from
scratch all the software needed to model a complex astrophysical system.
Whether a simulation requires a state-of-the-art stellar evolution code,
hydrodynamics code or stellar dynamics code, it has become standard
practice to use a `legacy code'.  One may write some modifications and
additions, but it is hard to compete with the tens of person-years
that have gone into some of the standard codes.

As a result, codes have steadily grown in complexity within separate
disciplines in astrophysics, but until recently little effort has been
invested to make these legacy codes compatible across different areas of
astrophysics.  In the area of dense stellar systems, the topic of this
meeting, the result has been that we have access to very detailed
codes that can model the evolution of individual stars, other codes
that can handle collisions between stars, and yet other codes that can
follow the motion of the whole ensemble of stars.  However, none of
these codes come even close to being able to talk with each other.

To let different codes talk together, similar requirements hold as for
humans trying to communicate between cultures and across language barriers:
each person needs to adapt to some extent to the customs of the other
culture, and there needs to be a dictionary to translate between the
languages.

The first requirement reads like a medical oath: {\it first of all, do
no harm}, \hbox{{\it i.e.} do not cause} the combined codes to halt.
It is better that a large-scale simulation is allowed to come to
completion, with messages in a log file indicating where serious
errors may have been made, than to have the simulation halt at each
bend of the road, when one of the modules is not optimally happy.

The second requirement boils down to the definition of specific
interfaces, and a willingness in the community to adopt a standard
definition.  Producing such a definition requires a fair amount of
care, flexibility and vision, to avoid the danger of saddling a field
with a standard that can hinder future progress.

In the next few pages, I summary recent efforts to let codes cross
the boundaries of stellar evolution, stellar dynamics and stellar
hydrodynamics, in order to enable next-generation large-scale
simulations of dense stellar systems.

\section{MODEST-1: New York}

It was the goal of the MODEST-1 meeting to begin addressing this
problem of letting codes talk to each other.  The workshop was held at
the Hayden Planetarium in the Museum for Natural History in New York
City, in June 2002.  The MODEST acronym was coined during this
meeting, and it can stand not only for MOdeling DEnse STellar systems,
but also for MODifying Existing STellar codes.  The latter description
stresses the desirability to start with what is already available, and
to find ways to put it all together, rather than to try to write a
kitchen-sink type over-arching super code from scratch.

The format of the meeting was unconventional.  The organizers, Piet
Hut and Mike Shara, polled whether there was enough interest to
warrant a workshop, and then sent out an email with the following
note: ``this will be a work-workshop, not a lectures-workshop.
Believe it or not, we have not a single scheduled lecture!  We are
curious to see how this format will work out.  Hopefully we will
actually get a lot of work done, together as well as in smaller
groups.''

We did.  The 34 participants met for five days, splitting up in
topical groups and reporting their results back to the group on a
daily basis.  Major progress was made in defining and starting to
tackle the two main requirements listed above, that of culture and
translation.  Within a few weeks of the meeting, 10 of the
participants had written a review paper, in lieu of proceedings, which
was posted on astro-ph, and subsequently published (Hut et al. 2003).

Also immediately following the workshop, a web site was established,
at \ {\tt http://www.manybody.org/modest.html}\ , as a place to
accumulate general information, demos, toy models, links to more detailed
models and simulation software, {\it etc.}  At the same time an email
list was started, with occasional announcements of major events and
progress reports.  The web site contains instructions for subscribing
to this email list.

As a result of the great enthusiasm displayed during the first
workshop, it was decided to hold similar meetings frequently enough so
as to enable ongoing projects to be critiqued at each meeting.  Such
peer-review feedback is essential for the health of any really
large-scale project, and therefore a frequency of two meetings per
year was chosen as an optimal compromise.  While individual projects
could of course use more frequent reviews, getting the main players
together every half year was already seen as quite a challenge.  It
was a challenge that was met, it turns out, judging from the
attendance of each of the following two workshops, between 30 and 40
people, comparable to that of MODEST-1.

\section{MODEST-2: Amsterdam}

MODEST-2 was organized by Simon Portegies Zwart en Piet Hut, at the
University ofu Amsterdam, Holland, in December 2002.  This workshop
followed a similar format: there were no scheduled talks.  Instead,
individual participants sometimes showed a couple view graphs, to
illustrate a particular point.  Also, during the first day a number of
specific topics of interest were formulated, and speakers were invited
to give an impromptu brief introduction to each topic.  In this way,
eight short talks were delivered by participants outlining how their
own work was fitting into the MODEST framework, what they wanted to
get out of participation in MODEST, what the most relevant questions
were for their area, and what they had accomplished since MODEST-1.

During the previous workshop in New York, a broad discussion had been
started concerning the question when to use full-blown stellar
evolution codes and when to use approximate recipes.  This discussion
was continued in Amsterdam.  The conclusions reached were similar, but
more refined in detail: to use recipes for single stars; a mix of
recipes and codes for interacting binaries; and `life' stellar
evolution codes for merger products.  Given the very many ways that
stars can form from single or even repeated mergers, there is just
no way that we can expect to construct a grid of model tracks to
anticipate the specific needs for evolving such merger products.  Not
only will the chemical compositions be different from standard values,
but incomplete mixing of the progenitor stars will add a full
functional amount of degrees of freedom.

Another outcome of this workshop was the organization of eight working
groups, listed below (the ninth working group was added during MODEST-3).
The main MODEST web site now contains pointers to these working groups.
While Steve McMillan is the webmaster for the main web site, the
following people are the contact persons for the web sites of the
individual working groups:

\medskip

\begin{tabular}{lll}
Working Group & & Contact Person\\
\hline
Star Formation & & Ralf Klessen\\
Stellar Evolution & & Onno Pols\\
Stellar Dynamics & & Rainer Spurzem\\
Stellar Collisions & & Marc Freitag\\
Simulating Observations of Simulations & & Simon Portegies Zwart\\
Data Structures & & Peter Teuben\\
Validation & & Douglas Heggie\\
Literature & & Melvyn Davies\\
Observations & & Giampaolo Piotto\\
\end{tabular}

\medskip

\noindent
The highlights of the workshop appeared on astro-ph within a few weeks
after concluding the meeting, and were published as a review
(Sills et al. 2003).

\section{MODEST-3: Melbourne}

The third workshop was held at Monash University in Melbourne,
Australia, and was organized by Rosemary Mardling and Piet Hut.  The
meeting was held in early July 2003, in the three days before
the General \hbox{Assembly assembled} \hbox{in Sydney.}  In contrast
to the previous two workshops, we experimented with a more traditional
style of scheduled presentations (see the MODEST-3 web 
\hbox{site \ \ {\tt http://www.manybody.org/modest/Workshops/modest-3.html}}
\ for the detailed program with speakers, titles, and abstracts).

These talks filled the first two days of the meeting.  In the morning
of the third day, each of the working groups reported about their
progress, and a ninth working group was added to focus on observations.
The afternoon featured an open discussion.  One of the outcomes was a
series of proposals for extending the web site, by including a page
for job opportunities and another page for project proposals.
Examples of the latter can be notes from people who have observational
data sets, for which they invite theoreticians to join them in
simulations of the systems observed; or simulators who have large data
sets for which they invite students to join them in analyzing the results.

After all the informal discussions during the first two workshops,
this meeting with more formal talks played a complementary role, as a
one-time occasion to inform people from all three fields (evolution,
dynamics, hydrodynamics) about the main activities in the other
fields.  Subsequent workshops will return to the old format of more
free-wheeling discussions, starting with MODEST-4, in January 2004
at Geneva Observatory, located between Geneva and Lausanne in
Switzerland.

\section{Future Meetings and Activities}

The MODEST initiative has grown in one year from an initial informal
meeting into a broad framework to facilitate collaborations across
boundaries between various disciplines within computational
astrophysics.  So far, these collaborations have taken on the form of
sharing codes, defining and building code interfaces, developing
demos, writing review papers, applying for joint grants, and starting
joint research projects.  Future workshops may coincide with the
organization of MODEST summer schools and other teaching activities,
as well as outreach projects aimed at the general public.  MODEST is
an open forum: we invite anyone interested in simulations of dense
stellar systems to join us.

\end{document}